\documentclass{svproc}
\usepackage{graphicx}%
\usepackage{subcaption}%
\usepackage{multirow}%
\usepackage{amsmath,amssymb,amsfonts}%
\usepackage{mathrsfs}%
\usepackage[title]{appendix}%
\usepackage{xcolor}%
\usepackage{textcomp}%
\usepackage{manyfoot}%
\usepackage{booktabs}%
\usepackage[labelfont=bf]{caption}%
\usepackage{algorithm}%
\usepackage{algorithmicx}%
\usepackage{algpseudocode}%
\usepackage{listings}%
\usepackage{tabularx}%
\usepackage{cite}
\usepackage[hyphens]{url}

\usepackage{wrapfig}

\usepackage[inline=true, margin=false]{fixme}
\usepackage{hyperref}

\renewcommand{\vec}[1]{\boldsymbol{\mathbf{#1}}}

\begin{document}
\mainmatter              
\title{Credit Attribution in Timestamped Soccer Networks Using Multivariate Hawkes Processes}
\titlerunning{Credit Attribution in Soccer Networks Using Multivariate Hawkes Processes}  
%
\author{Phillip W. Kornberg and Kevin S. Xu}
\authorrunning{Phillip W. Kornberg and Kevin S. Xu} 

\institute{Case Western Reserve University \\ 10900 Euclid Avenue, Cleveland, OH 44106-7071, USA\\
\email{\{pwk9,ksx2\}@case.edu}}

\maketitle              

\begin{abstract}
The two main traditional player statistics used in soccer are goals scored and assists, with an assist credited when a player’s pass immediately leads to a teammate scoring.
Attributing credit solely to the player who makes the final pass before a goal fails to account for the contributions of multiple players in the buildup to the goal. 
Consequently, traditional statistics provide only a limited view of player involvement. 
In this paper, we use multivariate Hawkes processes (MHPs) to model timestamped sequences of passes and shots, which form continuous-time networks. 
MHPs provide a branching mechanism where an event (a pass or shot) can be generated as a child of a previous event, which allows us to quantify each pass's contribution to a subsequent pass or shot. 
This approach distributes credit across multiple passes within an attacking sequence, making it applicable across a wide range of situations. 
Building on this framework, we propose expected goal involvements (xGI), a metric that measures a player's involvement in creating goal scoring opportunities based on their passes. 
This metric identifies players who play key roles in creating goal scoring opportunities even if they do not make the final pass, offering a more comprehensive measure of attacking involvement and player performance.

\end{abstract}

\section{Introduction}
Passing the ball between players on the same team is a crucial part of the sport of association football, also known as soccer. 
The collection of passes during a soccer match is naturally modeled as a continuous-time network with nodes denoting players and timestamped directed edges denoting passes from one player to a teammate at a specific time during the match. 
Such passing networks have been the topic of many prior studies \cite{gyarmati2014searching,buldu2018using,caicedo2020passing}, often with the goal of better understanding collective team performance and tactics. 
Shots on goal can also be modeled by treating the opposing team's goal as an additional node and adding an edge from the player taking the shot to the goal node. 

In this paper, we consider the problem of \emph{credit attribution} in soccer passing networks: for a given action in a soccer match, such as a specific pass or shot, what past actions contributed most to the given action? 
This is especially important when we want to assign credit for shots on goal. 
Traditionally, credit for a shot is given to two players, the player who shoots the ball (the shooter), and the player who passed the ball to the shooter (the passer). 
If the shot goes into the opposing team's goal, then the shooter is awarded a goal, and the passer is awarded an \emph{assist}, recognizing the important contribution of the passer.
If the shot does not go into the goal, then the passer does not receive an assist, despite the outcome of the shot being mostly out of the passer's control. 

Recent developments in soccer analytics have moved to separate the outcome of a shot from the actions leading up to the shot, resulting in the creation of the expected goals (xG) and expected assists (xA) metrics. 
The xA metric allows the player who passed the ball to the shooter to still receive credit for creating a goal scoring opportunity regardless of the outcome of the shot. 
However, no credit is passed further back, to the players who made the passes leading up to the final pass. 
These previous passes could still be very important contributions to the shot and should also receive some credit. 

\subsubsection{Key Contributions}
We make three main contributions in this paper. 
First, we propose to use multivariate Hawkes processes (MHPs) to model timestamped networks of passes and shots, providing a mechanism that can be used to estimate the \emph{responsibility} of each prior pass leading up to a pass or shot. 
Next, we use these responsibilities to create a new statistic, the \emph{expected goal involvements (xGI)}, that measures a player's involvement in creating goal scoring opportunities beyond the final pass based on their passing history. 
Finally, we apply these methods to analyze data from the 2015/2016 La Liga season at the team and league level.

\section{Background and Related Work}

\subsection{Soccer Analytics Metrics}
In recent years, soccer analytics has progressed significantly, moving away from simple cumulative statistics such as shots and total passes to more sophisticated metrics like expected goals (xG) along with expected assists (xA). Expected goals (xG) is a metric that assigns each shot a value between 0 and 1, representing the likelihood that a shot will result in a goal based on factors such as shot location and defensive pressure~\cite{lucey2014quality}. Closely related to expected goals is expected assists (xA), which assigns the expected goals value of the resulting shot to the player who made the final pass~\cite{footballanalyst_xa}. For example, if a player passes to a teammate who shoots, the xG value of that shot is credited to the passer in the form of an xA. 

While traditional xA provide a measure of a player's ability to create goal scoring opportunities, it is limited in scope because it only considers the pass immediately preceding a shot. This approach works well in situations where a single pass leads directly to a shot, but it becomes less effective in scenarios involving multiple passes in the buildup to a shot, which occur far more frequently in soccer. To address this limitation, metrics such as Opta's expected assists (xA)~\cite{Whitmore_xA_OptaAnalyst} estimate the likelihood that a pass will become an assist regardless of whether the receiver takes a shot. 
Another approach is to give credit equally to every player who participated in a possession. 
This is the approach taken by xGChain \cite{lawrence2018xgchain}, which assigns the xG of a shot in the possession to all of the players involved. 
A downside is that a player who makes the first pass in a long passing sequence receives equal credit to the player who makes the last pass, even though these events may be separated by multiple minutes and dozens of passes and may contribute unequally to the shot.

\subsection{Analyzing Passing Networks in Team Sports}
The study of soccer passing networks, where players are represented as nodes and passes as edges, has been widely used to identify key players within teams and characterize a team's play style \cite{caicedo2020passing}. 
Gyarmati and Kwak~\cite{gyarmati2014searching} use motifs in the passing networks to characterize a team's play style.
Clemente et al.~\cite{Clemente2020} use traditional network metrics, such as degree centrality, betweenness centrality, and closeness centrality, to identify key players and quantify their contributions. 
While these metrics provide insights into player involvement, they often fail to account for the quality or context of individual passes.

Another line of work focuses on building probabilistic models for timestamped event sequences in sports, including passing networks.
Xin et al.~\cite{xin2017continuous} propose a continuous-time network model for events in basketball, including passes. 
Grieshop et al.~\cite{grieshop2023continuous} propose a continuous-time stochastic process designed to model different aspects of a soccer match, including ball control, passing, and pace. 
These models have been used for exploratory analysis of different sports teams but have not been used for credit attribution.

\subsection{Multivariate Hawkes Processes and Credit Attribution}
Temporal point processes (TPPs) are often used to model timestamped events that are observed at irregular time intervals. 
TPPs serve as a natural model for events, such as passes and shots, in soccer and other sports. 
A TPP is characterized by its intensity function, which represents the instantaneous rate at which events occur over time. 
A Hawkes process \cite{laub2021elements} is a self-exciting TPP where the occurrence of an event may influence the occurrence of additional events shortly thereafter. 
The influence then decays over time according to a kernel function that is typically chosen to be an exponential function $\beta e^{-\beta(t)}$. 

A multivariate Hawkes process (MHP) models multiple variables that can mutually excite each other. 
This provides an interesting model for soccer events, as the occurrence of one pass can increase the intensity of a future pass or shot.
An exponential MHP is characterized by its conditional intensity function
\begin{equation*}
\label{eq:multivariate_hawkes}
    \lambda_i(t|\mathcal{H}_t) = \mu_i + \sum_{j=1}^d \sum _{t_l \in T_j, t_l<t} \alpha_{ij} \beta e^{-\beta(t-t_l)}, 
\end{equation*}
where $\mathcal{H}_t$ denotes the history (all events up to time $t$), $\mu_i$ denotes a baseline rate at which events in variable $i$ occur, $d$ denotes the number of variables in the MHP (the dimension), $T_j$ denotes the set of times that an event in variable $j$ occurs, $\beta$ denotes the rate of exponential decay of the influence of past events, and $\alpha_{ij}$ denotes the normalized jump size such that the product $\alpha_{ij} \beta$ corresponds to the increase in the intensity of variable $i$ triggered by an event in variable $j$.

In our proposed MHP model for soccer action sequences (see Section \ref{sec:proposed_model} for details), each variable $i$ corresponds to a specific \emph{pair of players}: a sender $i_s$ who initiates the pass and a receiver $i_r$. Therefore, $\alpha_{ij}$ measures the influence of a pass from $j_s \rightarrow j_r$ on a future pass from $i_s \rightarrow i_r$. 

An MHP can be used to compute the probability that an event $l$ is the parent of event $k$, often called the \emph{responsibility} of event $l$ for event $k$:
\begin{equation}
\label{eq:responsibility}
z_{kl} =
\begin{cases}
\frac{\alpha_{v(k)v(l)}\beta \exp[-\beta (t_k - t_l)]}{\mu_{v(k)} + \sum_{t_{l'} < t_k} \alpha_{v(k)v(l')} \beta \exp[-\beta (t_k - t_{l'})]}, & 0 < l < k, \\
\frac{\mu_{v(k)}}{\mu_{v(k)} + \sum_{t_{l'} < t_k} \alpha_{v(k)v(l')}\beta \exp[-\beta (t_k - t_{l'})]}, & l = 0,
\end{cases}
\end{equation}
where $v(k)$ denotes the variable in the MHP for event $k$, and $l = 0$ denotes that the event occurred from the baseline rate and is not a child of a previous event. 
These responsibilities can be used to attribute how much credit event $l$ should be given for event $k$. 

\subsubsection{Credit Attribution in Event Sequences}
The problem of attributing credit to multiple prior events is also encountered in online advertising, where the last ad that a user clicks on is often assigned all of the credit if the user then makes a purchase. 
This is often referred to as ``last-touch attribution,'' which is similar to the attribution provided by giving an assist to the player making the final pass before a goal. 
Attributing credit to multiple prior events is often referred to as multi-touch attribution, and different data-driven algorithms have been proposed for online advertising \cite{shao2011data,xue2026deep}.

Recent approaches have used MHPs to model soccer action sequences in order to assign different amounts of credit to different players. FeatHawkes \cite{verronneau2026feathawkes} represents each player as a dimension of an MHP. There is also a dimension for shots. Last-touch credit is computed using the branching matrix, equivalent to the excitation matrix $\vec{\alpha}$ for an exponential MHP, which captures how strongly one player's actions directly trigger another player's actions or a shot. 
To provide credit for previous touches, multi-touch credit is computed using the descendant matrix $\vec{D} = \vec{\alpha}(\vec{I} - \vec{\alpha})^{-1}$, which adds up both direct and indirect chains of influence to give each player's overall expected contribution. 

Another MHP-based approach is proposed by Baouan et al.~\cite{baouan2025generation}, who build a $12$-dimensional MHP representation with $11$ dimensions denoting the players' positions and a $12$th dimension denoting a ball touch in a predefined danger area around the opponent's goal. 
They similarly use the branching matrix $\vec{\alpha}$ and descendant matrix $\vec{D}$ for last-touch and multi-touch attribution, respectively, and use them to create direct and indirect generation of threat indices for players.

\section{Dataset Description and Preprocessing}

We use publicly available data from StatsBomb \cite{statsbomb2026statsbomb} and consider data from all of the teams in La Liga over the 2015/2016 season. 
To construct timestamped passing networks, we extract all pass and shot events from the raw JSON files separately for each half and team. Separating halves avoids complications from overlapping timestamps caused by stoppage time. 

\begin{wrapfigure}[20]{r}{1.5in}
\renewcommand{\tabcolsep}{3pt}
\centering
\vspace{-22pt}
\begin{tabular}[b]{ccc}
\hline
Time  & Sender & Receiver\\
\hline
10:00 & 4 & 0 \\
10:02 & 0 & 2 \\
10:05 & 2 & 8 \\
10:07 & 8 & 1 \\
10:09 & 1 & 0 \\
\hline
\end{tabular}
\\[6pt]
\includegraphics[width=1.5in]{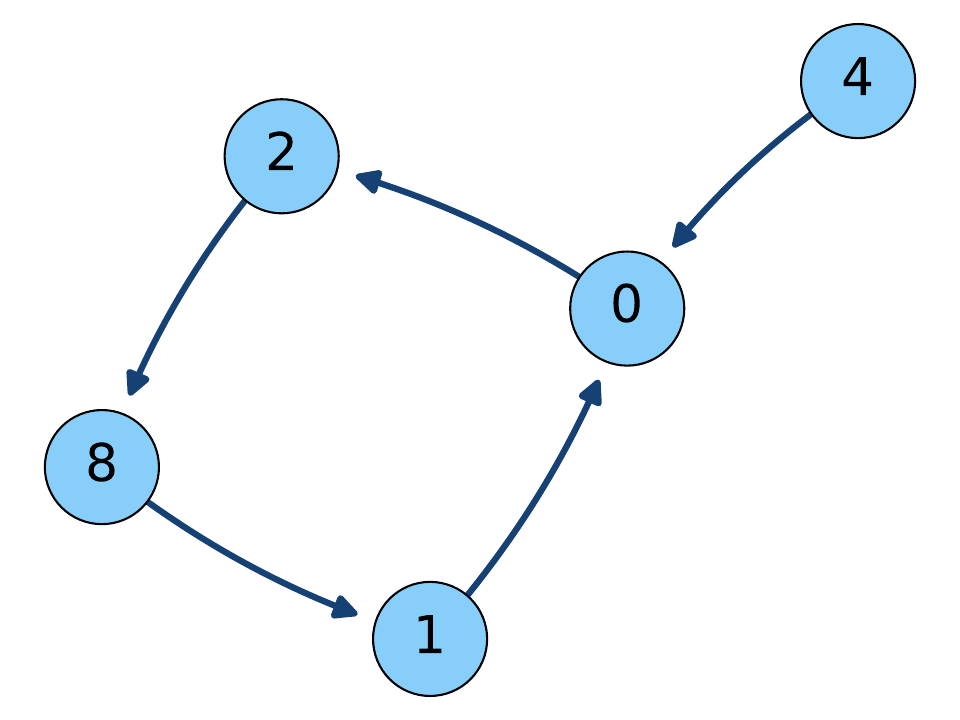}
\caption{A sequence of passes beginning with player 4 represented as a relational event table and directed graph.}
\label{fig:pass_seq}
\end{wrapfigure}

For passes, we record the sender, receiver, timestamp, and location using player IDs. For shots, we record the shooting player and expected goals (xG). The receiver for a shot is represented by a new goal node rather than the opposing goalkeeper, allowing shots to be treated as terminal actions rather than events that could influence future passes. Fig.~\ref{fig:pass_seq} displays a subset of one data frame and the corresponding directed graph. 

After constructing data frames for each half, we combine events from all matches in a season and store each half separately. A global player index is used to maintain consistent player mapping across matches. The data is then transformed from data frames into arrays, where each key represents a sender-receiver pair and the corresponding value is a list of timestamps when passes occurred between them. 
We add $600$ seconds to the times of all events in the second half to avoid the overlap with stoppage time events in the first half.
This process is repeated for each match, so the full dataset becomes a list of these arrays across the entire season.

\begin{figure}[tb]
    \centering
    \includegraphics[width=0.7\linewidth]{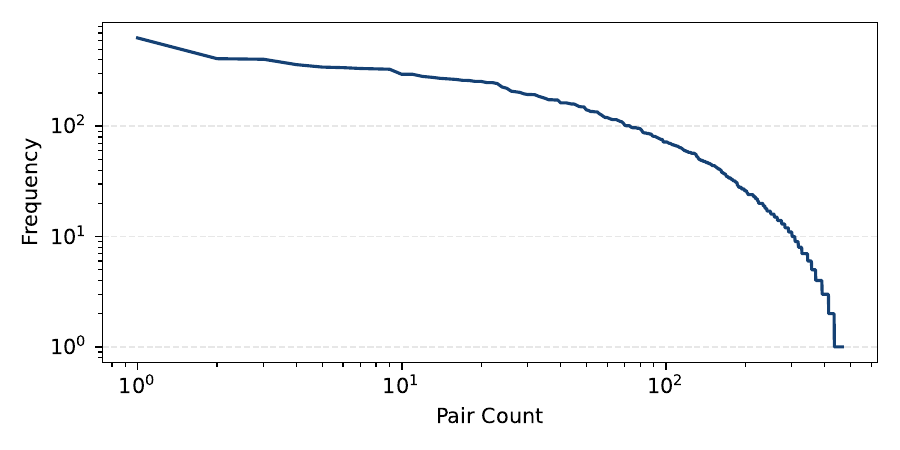}
    \caption{Frequency of passes for sender-receiver pair in Barcelona's 2015/2016 La Liga season, sorted in decreasing order.}
    \label{fig:pairCount}
\end{figure}

One challenge in transforming the data in this way was making sure that
every game uses the same set of player pairs. This is important because the MHP compares interaction patterns across matches, so the input structure must be consistent for each match. In our model, each team typically had around 25 players, which resulted in roughly 500 sender-receiver pairs. However, many of these pairs either occurred very rarely, as shown in Fig.~\ref{fig:pairCount}, or did not appear at all in certain matches. This creates both computational and statistical issues as rare interactions produce unstable alpha estimates. To address this issue, we only select the 100 most frequent pairs, which represent the most common and consistent passing interactions for each team.

\section{Methods}

\subsection{Multivariate Hawkes Process Model of Passing Networks}
\label{sec:proposed_model}
We model timestamped passing networks using an exponential multivariate Hawkes process (MHP), where each variable or dimension in the MHP corresponds not to a single player, unlike in \cite{baouan2025generation,verronneau2026feathawkes}, but to a \emph{pair of players}, which we denote as a \emph{sender-receiver pair}. 
As discussed previously, we include only the 100 most frequent sender-receiver pairs as well as pairs where the receiver is the goal node, so that all shots on goal are captured. 
Events involving other sender-receiver pairs are excluded from training. 

We use the HawkesTorch~\cite{raza2026hawkestorch} library, which allows for fast fitting of exponential MHPs using maximum likelihood. The model learns a full excitation matrix $\vec{\alpha}$, where entry $\alpha_{ij}$ captures how a pass between sender-receiver pair $j$ influences a pass between sender-receiver pair $i$ in the near future. 
Another model parameter is the vector of baseline rates $\vec{\mu}$, where $\mu_i$ models how frequently passes between sender-receiver pair $i$ occur in isolation when not caused by a prior pass. 
Finally, the scalar $\beta$ denotes the rate of exponential decay. 
To capture the fast paced nature of a soccer match where passes occur rapidly, we choose the decay parameter $\beta = 0.2$. 
With this choice, about $98\%$ of the influence of a pass on a future pass is gone after $20$ seconds.

Since fitting the MHP requires estimating a full excitation matrix, we evaluate several regularization techniques to prevent overfitting: $\ell_1$, $\ell_2$, and nuclear norm. 
We use Optuna \cite{akiba2019optuna} to select optimal penalty hyperparameters, which we compare using negative log-likelihood (NLL) on validation data. 
We normalize by the number of events to allow for comparison across matches and use a 70/15/15 train/validation/test split. We set the range from $10^{-8}$ to $10^{3}$ and run 1,000 trials. 
We find that the different penalty configurations have only small effects on the test NLL, with the unregularized model achieving the lowest value. 
Thus, we select the unregularized MHP model, which is the maximum-likelihood estimate (MLE). 
We train for 1,000 epochs, beyond which the NLL plateaus.

\subsubsection{Differences from Prior MHPs for Soccer Actions}
Rather than using players as dimensions or variables of the MHP as in \cite{baouan2025generation,verronneau2026feathawkes}, we use pairs of players. This helps preserve more structure than treating each player as one dimension, as it can capture patterns at the level of the passing network, not just individual player sequences. 
Furthermore, it allows us to directly perform multi-touch attribution through the estimated branching matrix $\vec{\alpha}$. 
This is unlike in \cite{baouan2025generation,verronneau2026feathawkes}, where $\vec{\alpha}$ provides only last-touch attribution, and the descendant matrix $\vec{D}$ is required to compute multi-touch attribution. 

\subsubsection{Smoothing Estimates}
A problem with the MLE is that $\alpha_{ij} = 0$ if sender-receiver pair $j$ never occurs shortly before a sender-receiver pair $i$ in the training data. 
This would result in $j$ receiving no credit for $i$, even if $j$ occurs one second before $i$ in the test data. 
Additionally, we may observe sender-receiver pairs that are not one of the 100 most frequent pairs and are thus not part of the model. 
These pairs would also receive no credit. 
From a soccer perspective, this is not realistic, as nearly any pass has at least some potential to contribute to a subsequent shot. To address this issue, we smooth the estimated $\vec{\alpha}$ by taking a convex combination of the MLE and a scaled all-ones matrix: 
\begin{equation*}
    \vec{\alpha} = (1-\epsilon)\vec{\alpha}^{\mathrm{MLE}} + \epsilon\left(\frac{0.5}{M}\right),
\end{equation*}
where $M$ is the total number of sender-receiver pairs. 
We choose $\epsilon=0.3$ so that the final $\alpha_{ij}$ value is 70\% based on the MLE and 30\% based on a small non-zero value $0.5/M$. 
The $0.5/M$ scaling is used to ensure the largest eigenvalue of the scaled all-ones is less than one, ensuring a stable MHP. 
Smoothing using a scaled all-ones matrix can be viewed as a uniform prior distribution and is similar to additive smoothing techniques commonly used in language modeling \cite{chen1999empirical}. 
The smoothing makes the effective floor of the matrix $\approx 0.001$ and ensures that an event involving pair $j$ occurring immediately before an event involving pair $i$ receives some credit, even if it is not observed in the training data.  
Additionally, if the sender-receiver pair $i$ or $j$ is not one of the 100 most frequent pairs, its $\alpha_{ij}$ value is the minimum value of $\vec{\alpha}$, the same as if $\alpha_{ij}^{\mathrm{MLE}} = 0$.

\subsection{Shot Responsibilities}
\label{sec:shot_responsibilities}
Using the value of $\beta$ along with the estimated $\vec{\alpha}$, $\vec{\mu}$, we can measure how much a pass contributed to a subsequent shot by computing the responsibility of the pass for the shot using \eqref{eq:responsibility}. 
For each shot, we examine the 6 preceding passes and calculate the probability that each pass contributed to the shot, along with the probability that the shot occurred independently of any pass, i.e., $l=0$ in \eqref{eq:responsibility}. Independently created or spontaneous shots are often a result of free kicks and penalty kicks, where there is a large time gap between passes and the shot. 

Importantly, the model does not assign responsibility solely based on the time gap between a pass and a shot. While the exponential decay with rate $\beta$ gives greater weight to more recent passes, the learned $\alpha_{ij}$ values also influence attribution. This allows passes that historically lead to subsequent shots more often to receive greater responsibility than more recent passes with lower $\alpha_{ij}$ values. Thus, the model accounts for both temporal proximity and the learned influence of each sender-receiver pair, providing a more nuanced attribution of credit than a fixed global $\alpha$ shared across all sender-receiver pairs, which would assign credit primarily on the time between events.

\subsection{Expected Goal Involvements (xGI)}
The shot responsibility values form the basis of a new metric we propose called expected goal involvements (xGI), which measures a player's involvement in creating goal scoring opportunities based on their passes. Rather than treating passes as independent actions or assigning all credit to the final pass, xGI distributes the expected goal (xG) value of a shot, which is part of the StatsBomb data \cite{statsbomb2026statsbomb}, among all players who contributed to the attacking sequence: 
\begin{equation*}
\text{xGI}_{l_s} = \sum_{k=1}^{m} \text{(xG)}_k \cdot z_{kl},
\end{equation*}
where $m$ denotes the number of shots in a game, and $l_s$ denotes the sender of pass $l$. 
This allows players who initiate or advance dangerous plays, but do not provide the final pass, to receive credit for their involvement.

\section{Results}

\subsection{Barcelona 2015/2016 Analysis}

\subsubsection{Model Fitting} 

\begin{figure}[b]
\centering
    \includegraphics[width=\linewidth]{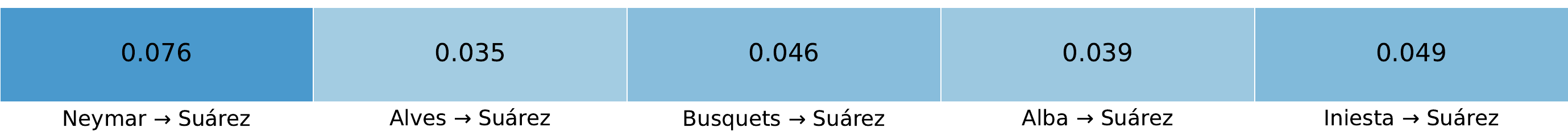}
    
    \vspace{1em}
    
    \includegraphics[width=\linewidth]{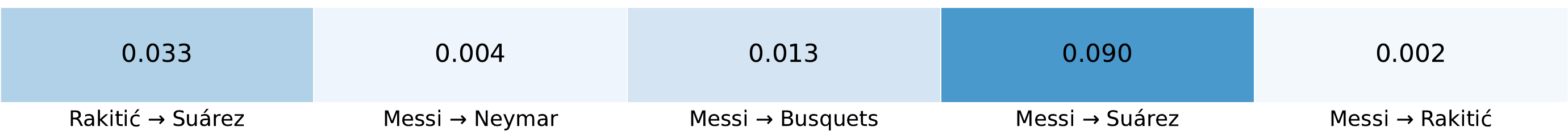}
\caption{Subset of $\alpha_{ij}$ values for passes leading up to a shot by Luis Suárez.}
\label{fig:goalAlphas}
\end{figure}

After filtering the data to include only the 100 most frequent pairs, we fit the MHP on a total of 17,645 pass and shot events. 
Fig.~\ref{fig:goalAlphas} depicts a subset of the estimated $\vec{\alpha}$ matrix, which shows the impact of passes on future shots by Luis Suárez.
Pairs such as (Neymar $\to$ Suárez) and (Messi $\to$ Suárez) have the largest $\alpha_{ij}$ values, which is expected since passes received by Suárez are more likely to directly contribute to shots taken by Suárez. However, the model assigns nonzero $\alpha_{ij}$ values to several pairs where the receiver is not Suárez, such as (Messi $\to$ Busquets). This indicates that passes not directly involving Suárez can still contribute indirectly to future shots by Suárez through subsequent passes, which allows for multi-touch attribution.

\subsubsection{Shot Responsibilities}
Using the entire $\vec{\alpha}$ matrix, we compute shot responsibilities for each individual match. Table \ref{tab:AllResponsibilities} shows two passing sequences alongside the responsibility values produced by our model. 
Notice that the differences in shot responsibilities depend both on the $\alpha_{ij}$ value and the time gaps between passes. 
For the shot in Table \ref{tab:evenly}, the responsibilities are distributed relatively evenly between the two passes leading up to the shot by Munir, as the penultimate pass (Suárez $\to$ Neymar) has the higher $\alpha_{ij}$ value than the final pass (Neymar $\to$ Munir). 
For the shot in Table \ref{tab:concentrated}, the pass (Piqué $\to$ Messi) is given by far the highest responsibility due to having both a much higher $\alpha_{ij}$ value compared to the other passes and being the final pass. 

\begin{table}[t]
\renewcommand{\tabcolsep}{6pt}
\centering
\caption{Examples of shots with (\subref{tab:evenly}) widely distributed and (\subref{tab:concentrated}) concentrated responsibilities from our proposed MHP model. Responsibilities from a temporal decay (TD) baseline model do not show much difference between the two shots.}
\label{tab:AllResponsibilities}
        \begin{subtable}{\textwidth}
        \centering
        \caption{Widely distributed shot responsibility (xG = 0.55)}
        \label{tab:evenly}
        \begin{tabular}{clcccc}
        \hline
        & & \multicolumn{2}{c}{Responsibilities} \\
        Time & Event & TD Baseline & MHP Model & xGI & $\alpha_{ij}$ \\
        \hline
            0:35 & Munir Shot & - & - & - \\
            0:34 & Neymar Pass to Munir & 0.473 & 0.329 & 0.1810 & 0.0013\\
            0:30 & Suárez Pass to Neymar & 0.213 & 0.268 & 0.1474 & 0.0023\\
            0:29 & Neymar Pass to Suárez & 0.174 & 0.136 & 0.0748 & 0.0014\\
            0:26 & Roberto Pass to Neymar & 0.096 & 0.086 & 0.0473 & 0.0016\\
            0:20 & Mathieu Pass to Roberto & 0.029 & 0.020 & 0.0110 & 0.0013\\
            0:17 & Bravo Pass to Mathieu & 0.015 & 0.011 & 0.0061 & 0.0013 \\
            - & Spontaneous Shot & - & 0.150 & - & - \\
        \hline
        \end{tabular}
\end{subtable}

    \vspace{1em}

    \begin{subtable}{\textwidth}
        \centering
        \caption{Concentrated shot responsibility (xG = 0.12)}
        \label{tab:concentrated}  
        \begin{tabular}{clcccc}
        \hline
        & & \multicolumn{2}{c}{Responsibilities} \\
        Time & Event & TD Baseline & MHP Model & xGI & $\alpha_{ij}$ \\
        \hline
            82:57 & Messi Shot & - & - & - \\
        82:56 & Piqué Pass to Messi & 0.559 & 0.649 & 0.0779 & 0.0057 \\
        82:52 & Rakitić Pass to Piqué & 0.251 & 0.075 & 0.0090 & 0.0015  \\
        82:48 & Busquets Pass to Rakitić & 0.112 & 0.031 & 0.0037 & 0.0014 \\
        82:45 & Alba Pass to Busquets & 0.062 & 0.018 & 0.0022 & 0.0014 \\
        82:36 & Messi Pass to Piqué & 0.010 & 0.003 & 0.0004 & 0.0013  \\
        82:33 & Rakitić Pass to Messi & 0.006 & 0.029 & 0.0035 & 0.0249 \\
        - & Spontaneous Shot & - & 0.195 & - & - \\

        \hline
        \end{tabular}
    \end{subtable}
\end{table}

\subsubsection{Comparison with Temporal Decay Baseline}

To test whether the learned excitation matrix $\vec{\alpha}$ provides meaningful attribution beyond temporal proximity alone, we compare our proposed MHP model with a simple temporal decay (TD) baseline. The baseline considers the same six passes preceding each shot and computes responsibilities purely based on an exponential temporal decay $e^{-\beta (t-t_l)}$. 
The responsibilities for the TD baseline are normalized to sum to 1 so that they are directly comparable to the MHP model, except that it does not have a baseline $\vec{\mu}$. 
Notice from Table \ref{tab:AllResponsibilities} that the TD model's responsibilities between the two shots are very similar, as it considers only the time gaps between the passes and the shot, unlike our proposed MHP model. 

\subsection{La Liga 2015/2016 Analysis}

We can look at the xGI of each player across the entire season to understand their expected involvement in creating goal scoring opportunities. Many of the players appearing in Table~\ref{tab:laLigaTopAvg} are well known for their ability to create chances. Lionel Messi and Neymar Junior, who have the highest xGI per game, consistently lead players in xA and key passes.

\begin{table}[t]
\setlength{\tabcolsep}{6pt}
\centering
\caption{Top 5 La Liga players by average xGI per game}
\label{tab:laLigaTopAvg}
    \begin{tabular}{llccc}
    \hline
    Player & Team & xGI & Games Played & xGI Per Game \\
    \hline
    Neymar Junior & Barcelona & 15.5 & 34 & 0.457 \\
    Lionel Messi & Barcelona & 12.4 & 33 & 0.376 \\
    Jorge Resurrección & Atlético Madrid & 11.1 & 35 & 0.316 \\
    Luis Suárez & Barcelona & 11.1 & 35 & 0.316 \\
    Manuel Nolito & Celta Vigo & 8.39 & 29 & 0.289 \\
    \hline
    \end{tabular}
\end{table}

\begin{figure}[tb]
    \centering
    \begin{subfigure}[t]{0.48\textwidth}
        \centering
        \includegraphics[width=\linewidth]{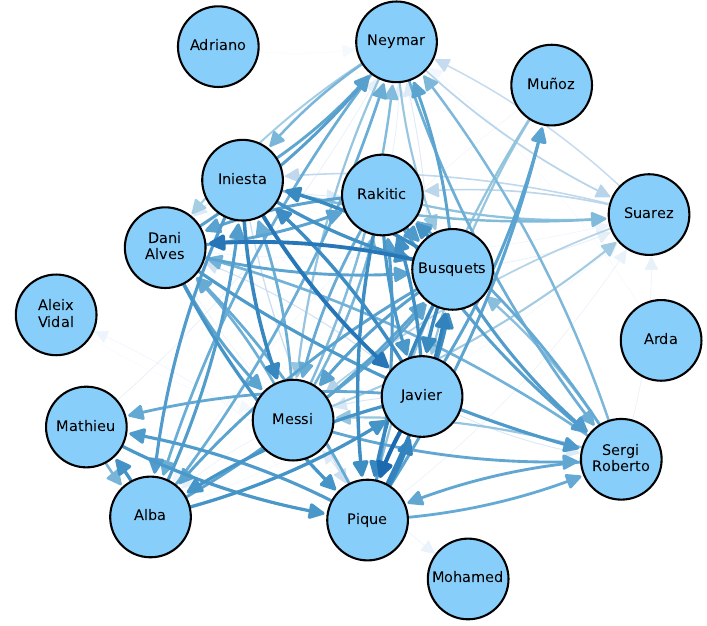}
        \caption{Barcelona}
        \label{fig:barcelona}
    \end{subfigure}
    \hfill
    \begin{subfigure}[t]{0.48\textwidth}
        \centering
        \includegraphics[width=\linewidth]{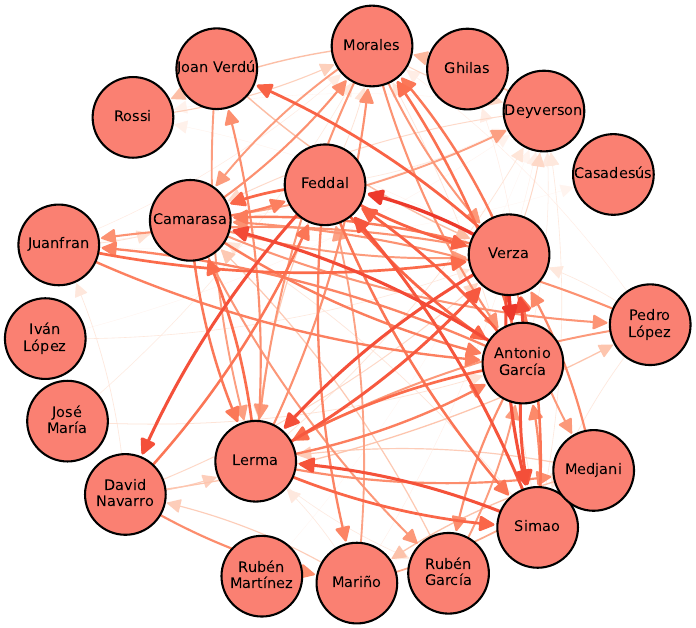}
        \caption{Levante UD}
        \label{fig:levante}
    \end{subfigure}
    
    \caption{2015/2016 La Liga influence networks for the 100 most frequent passing pairs. Edge width and color denote the total influence that a pass from the sender has on the receiver playing a pass shortly thereafter.
}
    \label{fig:PassingNetwork}
\end{figure}

Out of the 20 teams in La Liga, Barcelona has the highest mean $\alpha_{ij}$ value of 0.0050, while Levante UD has one of the lowest at 0.0040. 
Fig.~\ref{fig:PassingNetwork} shows the influence networks for different players on Barcelona and Levante UD. The weight of the edge for a sender-receiver pair $(s, r)$ is given by the sum of all $\alpha_{ij}$ entries for all $j$ corresponding to sender-receiver pairs $(s, :)$ and $i$ corresponding to sender-receiver pairs $(r, :)$, where the $:$ notation denotes all of the nodes. 
Unlike a passing network, where the edge weight measures the number of passes from a sender to a receiver, the influence network measures the total influence that the sender has on the receiver playing a pass in the future. 

From Fig.~\ref{fig:PassingNetwork}, Barcelona shows overall stronger influences between players compared to Levante UD. 
In particular, Barcelona's network has a stronger core, while Levante UD's network has more players on the periphery with only weak influences. 
This is consistent with the parameter estimates for the model.

\section{Conclusion}
This paper outlines a new approach to better characterize player involvement in creating goal scoring opportunities in soccer. Instead of attributing credit only to the final pass before a shot, our method distributes credit across multiple passes in the buildup. We accomplish this by modeling passing networks using an MHP and using a branching probabilities equation to compute shot responsibilities. Then, we introduce the expected goal involvements (xGI) statistic, which distributes xG based on player’s level of responsibility for creating the shot.

\subsubsection{Limitations}
Our model for credit attribution has several limitations. 
The first is the basic assumption---a pass ($s \rightarrow r$) that frequently leads to a pass ($s' \rightarrow r'$) shortly thereafter should be given more credit than a pass ($u \rightarrow v$) that infrequently leads to ($s' \rightarrow r'$). 
This method of attributing credit based on frequent occurrences from the training data may not be applicable to all application settings, and furthermore, the training data may not be sufficiently long to observe many such occurrences. 
Additionally, infrequent passes that are not among the top 100 most frequent pairs are currently excluded from the MHP model and are assigned the minimum $\alpha_{ij} \approx 0.001$. 
Both of these problems could potentially be alleviated by developing MHP-based network models that can group together different nodes, such as the multivariate community Hawkes (MULCH) model \cite{soliman2022multivariate}. Unfortunately, the MULCH model is not flexible enough in its current form for multi-touch attribution and can only perform last-touch attribution.

Another limitation of our analysis is the manual selection of $\beta$. 
Different $\beta$ values will result in a different $\vec{\alpha}$ matrix, and in turn, different shot responsibilities. Our analysis used $\beta = 0.2$, which we felt was a reasonable choice because it leads to $98\%$ of the influence of a pass on future passes being gone after $20$ seconds, which is a long time in a soccer match. However, $\beta$ could also reasonably be set to other values such as $0.1$ or $0.5$ to reflect different time scales.

\section{Acknowledgments}
This material is based upon work supported by the National Science Foundation grant IIS-2318751. 
This work was also supported by a summer research award provided by the CWRU Undergraduate Research Office.

\bibliographystyle{spmpsci} 
\bibliography{refs} 

@article{xin2017continuous,
  title={A CONTINUOUS-TIME STOCHASTIC BLOCK MODEL FOR BASKETBALL NETWORKS},
  author={Xin, Lu and Zhu, Mu and Chipman, Hugh},
  journal={Ann. Appl Stat.},
  pages={553--597},
  year={2017},
  publisher={JSTOR}
}

@misc{lawrence2018xgchain,
  author = {Lawrence, Thom},
  title = {Introducing {xGChain} and {xGBuildup}},
  year = {2018},
  month = {September},
  day = {30},
  organization = {Hudl},
  url = {https://www.hudl.com/blog/introducing-xgchain-and-xgbuildup},
  urldate = {2026-07-22}
}

@article{grieshop2023continuous,
  title={A Continuous-Time Stochastic Process for High-Resolution Network Data in Sports},
  author={Grieshop, Nicholas and Feng, Yong and Hu, Guanyu and Schweinberger, Michael},
  journal={arXiv preprint arXiv:2303.01318},
  url={https://arxiv.org/abs/2303.01318},
  year={2023}
}

@book{laub2021elements,
  author    = {Laub, Patrick J. and Lee, Young and Taimre, Thomas},
  title     = {The Elements of Hawkes Processes},
  publisher = {Springer International Publishing},
  address   = {Cham},
  year      = {2021}
}

@inproceedings{lucey2014quality,
  title={Quality vs Quantity: Improved Shot Prediction in Soccer Using Strategic Features from Spatiotemporal Data},
  author={Lucey, Patrick and Bialkowski, Alex and Monfort, Matthew and Carr, Peter and Matthews, Iain},
  booktitle={Proc. 9th Annu. MIT Sloan Sports Anal. Conf.},
  year={2015}
}

@misc{footballanalyst_xa,
  author    = {{The Football Analyst}},
  title     = {Expected Assists {(xA)} – Football Statistics Explained},
  year      = {2025},
  url       = {https://the-footballanalyst.com/expected-assists-xa-football-statistics-explained/}
}

@article{Clemente2020,
  title={Player position relationships with centrality in the passing network of world cup soccer teams: Win/loss match comparisons},
  author={Clemente, Filipe Manuel and Sarmento, Hugo and Aquino, Rodrigo},
  journal={Chaos, Solitons \& Fractals},
  volume={133},
  pages={109625},
  year={2020},
  publisher={Elsevier}
}

@article{raza2026hawkestorch,
  title = {Massively Parallel Exact Inference for {Hawkes} Processes},
  author = {Raza, Ahmer and Smith, Hudson},
  journal = {arXiv preprint arXiv:2604.01342},
  url = {https://arxiv.org/abs/2604.01342},
  year = 2026,
}

@misc{Whitmore_xA_OptaAnalyst,
  author       = {Whitmore, Jonny},
  title        = {What Are Expected Assists ({xA})?},
  year         = {2021},
  howpublished = {\url{https://theanalyst.com/articles/what-are-expected-assists-xa}}
}

@unpublished{verronneau2026feathawkes,
  title={The Path to the Goal: Feature-augmented {Hawkes} Processes for Event-level Attribution},
  author={Verronneau, Remi and Croissant, Lorenzo and Vieille, Bartholom{\'e} and Perchet, Vianney},
  year={2026},
  NOTE = {Preprint},
  url = {https://hal.science/hal-05284437}
}

@article{gyarmati2014searching,
  title={Searching for a unique style in soccer},
  author={Gyarmati, Laszlo and Kwak, Haewoon and Rodriguez, Pablo},
  journal={arXiv preprint arXiv:1409.0308},
  url={https://arxiv.org/abs/1409.0308},
  year={2014}
}

@article{buldu2018using,
  title={Using network science to analyse football passing networks: Dynamics, space, time, and the multilayer nature of the game},
  author={Buld{\'u}, Javier M and Busquets, Javier and Mart{\'\i}nez, Johann H and Herrera-Diestra, Jos{\'e} L and Echegoyen, Ignacio and Galeano, Javier and Luque, Jordi},
  journal={Front. Psychol.},
  volume={9},
  pages={1900},
  year={2018},
  publisher={Frontiers Media SA}
}

@article{caicedo2020passing,
  title={Passing networks and tactical action in football: A systematic review},
  author={Caicedo-Parada, Sergio and Lago-Pe{\~n}as, Carlos and Ortega-Toro, Enrique},
  journal={Int. J. Enviro. Res. Public Health},
  volume={17},
  number={18},
  pages={6649},
  year={2020},
  publisher={MDPI}
}

@inproceedings{akiba2019optuna,
  title={Optuna: A next-generation hyperparameter optimization framework},
  author={Akiba, Takuya and Sano, Shotaro and Yanase, Toshihiko and Ohta, Takeru and Koyama, Masanori},
  booktitle={Proc. 25th ACM SIGKDD Int. Conf. Knowl. Disc. Data Min.},
  pages={2623--2631},
  year={2019}
}

@article{chen1999empirical,
  title={An empirical study of smoothing techniques for language modeling},
  author={Chen, Stanley F and Goodman, Joshua},
  journal={Comput. Speech Lang.},
  volume={13},
  number={4},
  pages={359--394},
  year={1999},
  publisher={Elsevier}
}

@inproceedings{shao2011data,
  title={Data-driven multi-touch attribution models},
  author={Shao, Xuhui and Li, Lexin},
  booktitle={Proc. 17th ACM SIGKDD Int. Conf. Knowl. Discov. Data Min.},
  pages={258--264},
  year={2011}
}

@article{xue2026deep,
  title={Deep Learning for Multi-Touch Attribution in Online Advertising: A Comparative Study},
  author={Xue, Tao and Zhai, Panyu and Yang, Yanwu and Li, Yanling},
  journal={Inf. Syst. Front.},
  pages={1--35},
  year={2026},
  publisher={Springer}
}

@inproceedings{soliman2022multivariate,
  title={The {Multivariate Community Hawkes} model for dependent relational events in continuous-time networks},
  author={Soliman, Hadeel and Zhao, Lingfei and Huang, Zhipeng and Paul, Subhadeep and Xu, Kevin S},
  booktitle={Proc. 39th Int. Conf. Mach. Learn.},
  pages={20329--20346},
  year={2022}
}

@misc{statsbomb2026statsbomb,
  title={{StatsBomb} Open Data},
  author={{StatsBomb}},
  url={https://github.com/hudl/open-data},
  year={2026}
}

@article{baouan2025generation,
  title={Generation of threat: crediting football players for creating dangerous actions in an unbiased way},
  author={Baouan, Ali and Coustou, S{\'e}bastien and Lacome, Mathieu and Pulido, Sergio and Rosenbaum, Mathieu},
  journal={J. Quant. Anal. Sports},
  year={2025},
  publisher={De Gruyter}
}

\end{document}